\documentclass[fleqn,usenatbib]{mnras}
\usepackage{newtxtext,newtxmath}
\usepackage{epsfig}
\usepackage{physics}
\usepackage{amsmath, color, bm}
\usepackage{booktabs}
\usepackage{subfigure}

\usepackage{nameref}
\usepackage{graphicx}
\usepackage{graphics}
\usepackage[space]{grffile}
\usepackage{url}
\usepackage[utf8]{inputenc}
\usepackage{fancyref}
\usepackage{xcolor}
\usepackage{gensymb}
\usepackage[normalem]{ulem}
\usepackage{eqnarray,amsmath}
\usepackage{multirow}
\usepackage{multicol}
\usepackage{placeins}
\usepackage{float}

\title[Resolving the eccentricity of binary black holes]{Resolving the eccentricity of stellar mass binary black holes with next generation ground-based gravitational wave detectors}

\author[Pankaj Saini]{Pankaj Saini \thanks{E-mail: pankajsaini@cmi.ac.in}\\
Chennai Mathematical Institute (CMI), Siruseri, 603103, India\\
}

\date{Accepted XXX. Received YYY; in original form ZZZ}

\pubyear{2023}

\begin{document}
\label{firstpage}
\pagerange{\pageref{firstpage}--\pageref{lastpage}}
\maketitle

\begin{abstract}
Next generation ground-based gravitational wave (GW) detectors are expected to detect $\sim 10^4 \mbox{--} 10^5$ binary black holes (BBHs) per year. Understanding the formation pathways of these binaries is an open question. Orbital eccentricity can be used to distinguish between the formation channels of compact binaries, as different formation channels are expected to yield distinct eccentricity distributions. Due to the rapid decay of eccentricity caused by the emission of GWs, measuring smaller values of eccentricity poses a challenge for current GW detectors due to their limited sensitivity. In this study, we explore the potential of next generation GW detectors such as Voyager, Cosmic Explorer (CE), and Einstein Telescope (ET) to resolve the eccentricity of BBH systems. Considering a GWTC-3 like population of BBHs and assuming some fiducial eccentricity distributions as well as an astrophysically motivated eccentricity distribution (Zevin et $al.$ (2021)), we calculate the fraction of detected binaries that can be confidently distinguished as eccentric. We find that for Zevin eccentricity distribution, Voyager, CE, and ET can confidently measure the non-zero eccentricity for $\sim 3\%$, $9\%$, and $13\%$ of the detected BBHs, respectively. In addition to the fraction of resolvable eccentric binaries, our findings indicate that Voyager, CE, and ET require typical minimum eccentricities $\gtrsim 0.02$, $5\times 10^{-3}$, and $10^{-3}$ at $10$ Hz GW frequency, respectively, to identify a BBH system as eccentric. The better low-frequency sensitivity of ET significantly enhances its capacity to accurately measure eccentricity.
\end{abstract}




\section{Introduction}\label{sec:intro}
The LIGO \citep{LIGOScientific:2014pky} and Virgo \citep{VIRGO:2014yos} have detected $\mathcal{O}(100)$ compact binary mergers~\citep{LIGOScientific:2016aoc,LIGOScientific:2017vwq,LIGOScientific:2018mvr,LIGOScientific:2020aai,2020PhRvD.101h3030V,abbott2021gwtc,LIGOScientific:2020ibl,2021ApJ...922...76N,Olsen:2022pin,Nitz:2021zwj}. One of the most interesting questions these observations are capable of answering is related to the formation scenarios of compact binaries, especially of binary black holes (BBHs) which will be observed in large numbers with the present and next generation GW detectors. There are broadly two formation channels of compact binaries: {\it isolated formation channel} and {\it dynamical formation channel}~\citep{Mapelli:2018uds,Mapelli:2021taw,Mandel:2018hfr}. In isolated formation channel, two massive stars in a binary system co-evolve in the galactic field, undergo supernova explosions and form BBHs. In the dynamical formation, a BBH is formed due to gravitational interaction in dense stellar environments such as globular and nuclear star clusters. The current observations of BBH mergers by LIGO and Virgo pose a challenge in terms of being explained solely through a single formation channel. A mixture of formation channels is preferred over a single formation channel~\citep{LIGOScientific:2020kqk,Bouffanais:2021wcr,PhysRevD.103.023026,Zevin:2020gbd}.

Future gravitational wave (GW) detectors with improved sensitivity will detect a large number of BBH mergers. For example, a LIGO detector with Voyager technology is expected to detect $\sim 10^{4}$ BBH mergers per year~\citep{Baibhav:2019gxm}. Moreover, third-generation (3G) detectors such as Cosmic Explorer (CE)~\citep{LIGOScientific:2016wof,Reitze:2019iox} and Einstein Telescope (ET)~\citep{Punturo:2010zz,Sathyaprakash:2011bh,Maggiore:2019uih} are expected to detect $\sim 10^5$ BBH mergers per year~\citep{Baibhav:2019gxm,evans2023cosmic,gupta2023characterizing}. Different formation channels leave unique imprints on the properties of BBH systems, providing valuable clues about their formation history.

Orbital eccentricity is one such unique feature that can give us clues about the formation channels of the binaries as different formation channels are expected to follow different eccentricity distributions. The ellipticity of a binary's orbit at a given time depends on how these binaries formed. Since the emission of GWs carries away the energy and angular momentum from the binary system, with time, the elliptical orbits tend to become circular~\citep{PhysRev.136.B1224}. In the small eccentricity limit, the eccentricity approximately decreases with GW frequency as $e_t/e_0 \approx (f_0/f)^{19/18}$~\citep{PhysRev.136.B1224}. Here $e_t$ is the `time-eccentricity' in the {\it quasi-Keplerian} representation of the binary orbit~\citep{Damour:2004bz} when binary emits the GWs at dominant mode frequency $f$, and $e_0$ is the initial value of $e_t$ at a reference frequency $f_0$. \footnote{The definition of eccentricity is model dependent. One should ideally use a model-agnostic definition of eccentricity when comparing eccentricity measurements from different analyses~\citep{Bonino:2022hkj,2023arXiv230211257A}.} For example, a binary with an initial orbital eccentricity $e_0=0.6$ when emitting GWs at $0.1$ Hz reduces to one with eccentricity $\sim 0.1$ when it enters the ET band at $1$ Hz. The eccentricity reduces to $\sim 0.02$ when it starts emitting GWs in the frequency band of CE and Voyager at $5$ Hz. The eccentricity further reduces to $\sim 5\times10^{-3}$ when it emits GWs at $20$ Hz (the lower cut-off frequency of LIGO and Virgo).

In current GW detectors, the measurement of eccentricity for GW150914-like BBHs requires $e_0\gtrsim 0.05$ (at $10$ Hz GW frequency)~\citep{PhysRevD.98.083028,Favata:2021vhw}. Next generation GW detectors with improved sensitivity will enable us to constrain smaller eccentricities. Apart from the overall improvement in sensitivity, 3G detectors will have excellent low-frequency sensitivity where binaries can retain a larger eccentricity. LIGO-Virgo-KAGRA currently employ quasi-circular templates for the detection of BBHs. Template-based matched-filter searches for eccentric binaries require accurate eccentric waveform model. Including eccentricity in parameter space increases its dimensionality and hence the number of templates and computational cost. Various alternate search methods have been proposed for the detection of BBHs~\citep{Tiwari:2015gal, Cheeseboro:2021rey, Lenon:2021zac, 2023arXiv230703736P}.

Reference~\citep{Lenon:2020oza} reanalysed the detected binary neutron star (BNS) signals GW170817~\citep{LIGOScientific:2017vwq} and GW190425~\citep{LIGOScientific:2020aai} with eccentric waveform models and found that these systems have $e_0\leq0.024$ and $e_0\leq0.048$, respectively. Reference~\citep{Chen:2020lzc} studied the signal-to-noise ratio (SNR) sky distributions of eccentric and quasi-circular BBHs for second and third-generation GW detectors. The potential of next generation GW detectors to measure orbital eccentricity remains unexplored. {\it In this paper, we study the ability of next generation GW detectors to constrain the eccentricity of BBH systems once a BBH system has been confidently detected}.

Typically, binaries formed through isolated formation channels are expected to be in quasi-circular orbits by the time they enter the frequency band of ground-based detectors~\citep{PhysRev.136.B1224, 2008PhRvD..77h1502H}. Isolated binaries can have a non-negligible eccentricity due to the large natal kick imparted due to the second supernova. However, it happens well before the merger that the binary is likely to shed away all its eccentricity by the time it is observed in the frequency band of the ground-based detector. N-body simulations suggest that the formation of highly eccentric binaries is inevitable in dense stellar environments~\citep{Antonini:2012ad,Antonini:2013tea,Antonini:2015zsa,PhysRevLett.120.151101}. Dynamically formed binaries in dense star clusters such as globular star clusters, and nuclear star clusters can retain moderate to high values of eccentricity when observed in the frequency band of ground-based detectors~\citep{Wen:2002km,OLeary:2008myb,Bae:2013fna,Samsing:2013kua,2017ApJ...846...82Z,PhysRevD.97.103014,Gondan:2017wzd,PhysRevLett.120.151101,Rasskazov:2019gjw,Zevin:2018kzq,Grobner:2020drr,Gondan:2020svr,Tagawa:2020jnc,Zevin:2021rtf,2023arXiv230307421D}.

Close to the centre of the supermassive black hole (SMBH), stellar mass binaries can be bound to the SMBH and form a triple system. Highly eccentric binaries can be formed through this secular GW evolution~\citep{2010ApJ...713...90A, Antonini:2012ad, 2013MNRAS.431.2155N,  2014ApJ...794..122M, 2015ApJ...799..118P, 2017ApJ...841...77A,2018ApJ...856..140H, Hamers:2021kby}. The gravitational interaction of a binary system with the third body can also increase its eccentricity~\citep{Silsbee:2016djf,Liu:2018nrf,2018ApJ...863....7R}. The potential mechanisms in triple star systems that can lead to the formation of high eccentricity include isolated stellar flyby interaction of wide stellar mass BBHs in the field~\citep{Michaely:2019aet,Raveh:2022ste}, wide binaries in the galactic centre~\citep{2023arXiv231002558M, 2023ApJ...955..134R}, evolution of wide hierarchical triple systems under the influence of galactic tidal field~\citep{Grishin:2021hcp}, mergers in gaseous environment~\citep{2012MNRAS.425..460M, PhysRevLett.120.261101, 2022ApJ...931..149R}, dynamical instability induced due to mass-loss or mass transfer between stars~\citep{2012ApJ...760...99P}. Around $5\%$ of all dynamical mergers in globular clusters can give rise to BBHs with eccentricities $\gtrsim0.1$ at $10$ Hz GW frequency~\citep{2017ApJ...840L..14S,2018MNRAS.481.5445S,PhysRevD.97.103014,2018PhRvD..98l3005R,PhysRevLett.120.151101}.~\footnote{The unmodelled binary eccentricity will lead to the biased parameter estimation of compact binaries [see for e.g. \citep{Favata:2013rwa,Favata:2021vhw,PhysRevD.106.084031,PhysRevD.107.024009, 2023arXiv231108033S}].}

Next generation ground-based detectors would observe compact binaries at frequencies $\gtrsim 1$ Hz. Since binary is expected to be more eccentric in the past, detectors observing in the low-frequency regime would be able to constrain eccentricity with high accuracy. The Laser Interferometer Space Antenna (LISA)~\citep{Babak:2017tow} is a planned space-based mission that will be observing in the frequency range $10^{-4}-0.1$ Hz. Mergers of SMBHs will be the main targets of LISA. LISA can measure eccentricities $e_0>10^{-2.5}$ (at one year before the merger) for supermassive BBHs~\citep{2023MNRAS.tmp.3335G}. However, there is a possibility of observing heavy stellar mass BBHs by LISA during their early inspiral before merging in the frequency band of ground-based detectors~\citep{Sesana:2016ljz,Vitale:2016rfr,Moore:2019pke,Wong:2018uwb,Ewing:2020brd}. The eccentricity measurement of stellar mass BBHs in the LISA band can be used to distinguish between formation channel of these binaries~\citep{PhysRevD.94.064020,2019arXiv190702283R}. Combining the information from LISA and high-frequency observations of next generation ground-based detectors such as CE and ET can improve the parameter estimation of compact binaries. The multiband observation of stellar-mass black hole binaries can help in better measurement of eccentricity~\citep{2022arXiv220403423K}.

There are already claims for the presence of eccentricity in the detected GW events~\citep{Romero-Shaw:2020thy,Romero-Shaw:2021ual,2021arXiv210707981O,Gayathri:2020coq,Romero-Shaw:2022xko}. However, the eccentricity can be mimicked by spin-precession effects. Due to the lack of an inspiral-merger-ringdown waveform model that includes both eccentricity and spin precession, it is hard to distinguish between the two effects, especially for the short signals such as GW190521~\citep{Romero-Shaw:2022fbf}. However, eccentricity is not mimicked by the spin-induced precession for long-duration signals~\citep{2023arXiv230916638D}.

Apart from the eccentricity, the spin orientation of the component BHs in a binary can also be used to discern between different formation channels~\citep{Rodriguez:2016vmx,Stevenson:2017dlk,PhysRevD.96.023012,Vitale:2015tea,Farr:2017uvj,2017PhRvL.119a1101O,Farr:2017gtv,PhysRevD.98.084036,2018PhRvD..98h4036G}. Due to tidal interactions, isolated binaries are expected to have their spins aligned with the orbital angular momentum of the binary. However, dynamically formed BBHs are likely to have their spins misaligned. Hence these BBHs are expected to have isotropic spin distribution~\citep{2016ApJ...832L...2R}. Redshift evolution of merger rate can also be used to constrain formation scenarios~\citep{2018ApJ...866L...5R,2018ApJ...863L..41F,Ng:2021sqn}.

Here, we focus on the measurement of eccentricity to distinguish circular and eccentric binary systems. If the eccentricity of a BBH system is measured with relative error $\Delta e_0/e_0<1$, we will call it a {\it resolved} binary. This study examines the ability of Voyager, CE, and ET to measure the eccentricity of BBHs, drawn from a particular eccentricity distribution. More precisely, considering few fiducial eccentricity distributions as well as an astrophysically motivated eccentricity distribution given in~\cite{Zevin:2021rtf}, we quantify the fraction of detected binaries that can be confidently identified as eccentric. We also find that it requires typical eccentricities $\gtrsim 0.02, 5\times 10^{-3}, 10^{-3}$ at $10$ Hz GW frequency for Voyager, CE, and ET, respectively, to distinguish between eccentric and circular binary systems. In Section~\ref{sec:detectors}, we discuss the GW detectors considered in the study. Section~\ref{sec:bbh population} discusses the simulated binary black hole population. Section~\ref{sec:Methodology} explains the method of calculating statistical errors using Fisher information matrix and the eccentric waveform model used for the analysis. Section~\ref{sec:results} describes the results of the paper. Section~\ref{sec:summary} provides the summary and outlook of the study. Throughout the paper, we use units $G=c=1$.

\section{Detectors, Binary black hole population, and Methodology}\label{sec:details of bbh population}
\subsection{Detectors}\label{sec:detectors}
We consider three ground-based GW detector configurations:
\begin{enumerate}
    \item A LIGO detector operating at the Voyager sensitivity~\citep{LIGO:2020xsf} with lower cut-off frequency $f_{\rm low}=5$ Hz. \vspace{2mm}
    \item Cosmic Explorer~\citep{Reitze:2019iox} in USA with $40$ km arm length at its design sensitivity. The lower cut-off frequency is chosen to be $f_{\rm low}=5$ Hz. \vspace{2mm}
    \item Einstein Telescope~\citep{Punturo:2010zz} in Europe with $10$ km arm length at its design sensitivity. This will be an underground facility with a triangular shape. The lower cut-off frequency is taken to be $f_{\rm low} = 1$ Hz. 
\end{enumerate}

\subsection{Binary black hole population}\label{sec:bbh population}
We consider a GWTC-3 like population of stellar mass BBHs with the following distributions:
\begin{itemize}
    \item {\it Primary mass:} The primary mass ($m_1$) of BBH population follows {\tt PowerLaw+Peak} model~\citep{GWTC3-population} which is a mixture of a power law and a Gaussian distribution~\citep{Talbot:2017yur}. The values of parameters defining the {\tt PowerLaw+Peak} model are taken to be: $\lambda_{\rm peak}=0.04$, $\alpha=3.4$, $m_{\rm min} = 5.08$, $m_{\rm max}=86.85$, $\mu_m=33.73$, $\sigma_m = 3.56$, $\delta_m=4.83$. \vspace{3mm}
    \item {\it Mass ratio:} The mass ratio of BBH population follows a power law distribution $p(q) = q^{\beta}$ with $\beta=1.08$. \vspace{3mm}
    \item {\it Spins:} We assume that the spin angular momentum vectors of binary components are aligned with the orbital angular momentum vector of the binary (non-precessing). The magnitudes of dimensionless spins for both black holes are drawn from a {\tt Beta} distribution~\citep{Talbot:2017yur}
    \begin{equation}
    p(\chi_{1,2}|\alpha_\chi, \beta_\chi) \propto \chi_{1,2}^{\alpha_\chi -1} (1-\chi_{1,2})^{\beta_\chi-1}\,,
\end{equation}
where $\alpha_\chi$ and $\beta_\chi$ are the shape parameters that determine the mean and variance of the distribution. The values are $\alpha_\chi= 1.6$, $\beta_\chi=4.11$. \vspace{3mm}
\item {\it Eccentricity distributions:}
Table~\ref{tab:table_eccentricity} summarizes the eccentricity distributions considered in this study. We consider two fiducial eccentricity distributions: {\tt Loguniform} and {\tt Powerlaw} with two eccentricity ranges: $e_0 \in(10^{-4}, 0.2)$ and $e_0 \in (10^{-7}, 0.2)$. In Powerlaw distribution, the powerlaw index $1.2$ is chosen such that it does not deviate significantly from the Loguniform distribution. The upper limit of distributions is fixed to be $e_0=0.2$, beyond which the waveform model employed in this study becomes less accurate. We also consider an expected astrophysical eccentricity distribution based on cluster simulations given in Fig.~1 of~\cite{Zevin:2021rtf}. This will be referred to as {\tt Zevin} eccentricity distribution from here onwards. The Fig.~1 of ~\cite{Zevin:2021rtf} contains two eccentricity distributions plotted by dashed and solid lines. We use the eccentricity distribution shown in dashed lines which represents the intrinsic eccentricity distribution of the considered population. Similar to the Loguniform and Powerlaw eccentricity distributions, we restrict Zevin eccentricity distribution to the range $e_0 \in (10^{-7},0.2)$. Note that Zevin eccentricity distribution is based on cluster simulations which are based on many astrophysical assumptions. We draw eccentricity samples from the one-dimensional marginalized distribution and do not account for the correlations of eccentricity with other binary parameters, assuming them to be small.
\begin{table}
\centering
\renewcommand{\arraystretch}{1.8}
\setlength{\tabcolsep}{4pt}
\caption{Eccentricity distributions considered in this study. The initial eccentricity $e_0$ is defined at $10$ Hz GW frequency.}
\begin{tabular}{ccccc}
\toprule
\toprule
Eccentricity distribution & \textbf{Loguniform} & \textbf{Powerlaw} & \textbf{Zevin} \\
\midrule
PDF & $\rm{p}(\rm{e}_0) \propto \rm{e}_0^{-1}$ & $\rm{p}(\rm{e}_0) \propto \rm{e}_{0}^{-1.2}$    \\

\multirow{2}{*}{Range of $e_0$} &  $(10^{-4}-0.2)$ & $(10^{-4}-0.2)$ & \multirow{2}{*}{$(10^{-7}-0.2)$} \\
& $(10^{-7}-0.2)$ & $(10^{-7}-0.2)$   \\
\bottomrule
\bottomrule
\end{tabular}
\label{tab:table_eccentricity}
\end{table} \vspace{3mm}
    \item {\it Redshift:} Sources are distributed uniformly in comoving volume according to the following redshift distribution
\begin{equation}
    p(z) \propto \frac{1}{1+z} \frac{dV_c}{dz}\,.
\end{equation}
where $\frac{dV_c}{dz}$ is the comoving volume element at redshift z. The factor $(1+z)^{-1}$ converts the detector frame time to the source frame time. We choose maximum redshift $z_{\rm max}=2.5$ which corresponds to a luminosity distance $d_L = 20.9$ Gpc. The corresponding $d_L$ is calculated assuming {\tt Planck18}~\citep{Planck:2018vyg} cosmology using {\tt Astropy}~\citep{2018AJ....156..123A}.
\end{itemize} 

Note that there might be correlations of eccentricity distribution with other binary parameters such as mass, mass-ratio, and redshift distribution in an astrophysical N-body dynamical model (Zevin distribution) and one should in principle use multiparameter distribution for drawing population. The measurement errors in eccentricity however strongly depend on the eccentricity itself and weakly dependent on other binary parameters. This can be seen from the approximate leading-order scaling in Eq.~(4.18b) in ~\cite{Favata:2021vhw}. The measurement errors in eccentricity for a given detector scale as the inverse square of eccentricity. Therefore small correlations of eccentricity with other binary parameters are less likely to affect our results.

\subsection{Methodology}\label{sec:Methodology}
Using the BBH population discussed in Sec.~\ref{sec:bbh population}, we calculate measurement uncertainties on binary parameters. The statistical errors on binary parameters can be forecasted using the {\it Fisher information matrix} framework~\citep{Finn:1992wt,Cutler:1994ys,Poisson:1995ef}. This framework gives the $1\sigma$ width around the injected values of the binary parameters.
The inner product between two frequency-domain signals ${\tilde h}_1(f)$ and ${\tilde h}_2(f)$ is defined as 
\begin{equation}\label{eq:innerproduct}
    (h_{1}|h_{2}) = 2 \int_{f_{\rm low}}^{f_{\rm high}}\frac{\Tilde{h}_{1}^{*}(f) \Tilde{h}_{2}(f) + \Tilde{h}_{1}(f)\Tilde{h}_{2}^{*}(f) }{S_{n}(f)}\, df,
\end{equation}
where $S_{n}(f)$ is the one-sided noise power spectral density (PSD) of the detector and $\ast$ represents the complex conjugation. The limits of integration in Eq.~\eqref{eq:innerproduct} are fixed by the sensitivity of the detector and the properties of the source. The SNR $\rho$ is defined as the norm of the signal,
\begin{equation}\label{eq:snr}
    \rho^{2} = (h|h) = 4\int_{f_{\rm low}}^{f_{\rm high}}\frac{|\Tilde{h}(f)|^{2}}{S_{n}(f)}\,df\,,
\end{equation}
The Fisher information matrix $\Gamma_{ab}$ is defined as
\begin{equation}\label{fisher}
    \Gamma_{ab} = \Bigg(\frac{\partial h}{\partial \theta_{a}}\Bigg|\frac{\partial h}{\partial \theta_{b}}\Bigg)\,.
\end{equation}
where $\theta_{a}$ is the set of waveform parameters. The covariance matrix is obtained by taking the inverse of the Fisher matrix 
\begin{equation}\label{covariance}
 \Sigma_{ab}=\Gamma_{ab}^{-1}\;,   
\end{equation}
1$\sigma$ errors on binary parameters $\theta_{a}$ are calculated by taking the square root of the diagonal elements of the covariance matrix
\begin{equation}\label{error}
    \Delta \theta_a = \sqrt{\Sigma_{aa}} \;.
\end{equation}
\begin{table}
\centering
\renewcommand{\arraystretch}{1.2}
\caption{Fraction of binaries that are detected with $\text{SNR}\geq8$. The median value of five independent realisations is quoted. The median value is rounded to its nearest integer value.}
\begin{tabular}{ccc}
  \toprule
  \toprule
  Detector & Fraction of detected binaries (\%) \\
  \midrule
    \textbf{Voyager}  & $33$ \\
    \textbf{CE} &   $100$    \\
  \textbf{ET} &    $97$  \\
  \bottomrule
  \bottomrule
\end{tabular} 
\label{tab:snr}
\end{table}
\begin{table*}
\centering
\renewcommand{\arraystretch}{1.2}
\begin{tabular}{cccc}
\toprule
\toprule
& \multicolumn{3}{c}{Fraction (in \%) of detected binaries with $\Delta e_0/e_0<1$} \\
\midrule
Eccentricity distribution & \textbf{Voyager}  & \hspace{8mm}\textbf{CE} & \textbf{ET} \\
\midrule
\textbf{Loguniform$(\bm{10^{-4},0.2)}$}  & $14$ &\hspace{8mm}\ $36$  & $55$\\
\textbf{Loguniform}$(\bm{10^{-7},0.2)}$ & $8$ &\hspace{8mm}\ $20$ & $30$\\
\textbf{Powerlaw}$\bm{(10^{-4},0.2)}$  & $7$ &\hspace{8mm}\ $20$ & $37$\\
\textbf{Powerlaw}$\bm{(10^{-7},0.2)}$  & $2$ &\hspace{8mm}\ $5$ & $9$\\
\textbf{Zevin}$\bm{(10^{-7},0.2)}$ & $3$ &\hspace{8mm}\ $9$ & $13$\\
\bottomrule
\bottomrule
\end{tabular}
\caption{The fraction of detected binaries that are measured with $\Delta e_0/e_0<1$ for Voyager, CE, and ET. To account for statistical fluctuations of the results, the exercise is carried out five times and the median values are reported.}
\label{tab:fraction}
\end{table*}

We use {\tt TaylorF2Ecc} waveform model~\citep{Moore:2016qxz} which is an inspiral waveform model and accounts for the leading order $[\mathcal{O}(e_0^2)]$ corrections due to eccentricity in the phasing expression. The $\tt TaylorF2Ecc$ waveform model is valid for small eccentricities $e_0\lesssim 0.2$. The waveform model does not account for the eccentricity corrections to the amplitude. Since GW detectors are more sensitive to the GW phase than to the amplitude, small eccentricity corrections to the amplitude will be less important than eccentricity corrections to the phase. The full expression for GW phasing can be found in Eq.~(6.26) of \cite{Moore:2016qxz}. The circular part of Eq.~(6.26) in~\citep{Moore:2016qxz} is for non-spinning binaries. We add aligned-spin terms to the circular part of Eq.~(6.26)~\citep{Arun:2004hn,Arun:2008kb,Buonanno:2009zt,Mishra:2016whh}.

The lower limit of integration ($f_{\rm low}$) in Eq.~\eqref{eq:innerproduct} is set by the lower cut-off frequency of the detector. For Voyager\footnote{The noise curve for Voyager is taken to be voyager\_cb.txt file from \url{https://dcc.cosmicexplorer.org/public/0163/T2000007/005/} inside ce\_curves.zip.} and CE, $f_{\rm low}$ is set to be $5$ Hz. For ET, $f_{\rm low}$ is taken to be $1$ Hz. The upper cut-off frequency ($f_{\rm high}$) is the redshifted frequency corresponding to the innermost stable circular orbit $(f_{\rm isco})$ of the remnant Kerr BH~\citep{1972ApJ...178..347B,Husa:2015iqa,Hofmann:2016yih}. The $f_{\rm isco}$ is a function of two-component masses and spins. The full expression can be found in appendix C of \cite{Favata:2021vhw}. For CE, we use the noise PSD from~\cite{2017CQGra..34d4001A}. The analytical fit for CE noise PSD is given in Eq.~(3.7) of Ref.~\cite{Kastha:2018bcr}. For ET, we use design sensitivity ET-D~\citep{Hild:2010id}. The ET noise PSD is scaled by a factor of $(\sin^260^\circ)^{-1}$ to convert it to the effective sensitivity of ET's triangular design.

\section{Results}\label{sec:results}
In this section, we examine the capability of Voyager, CE, and ET to measure the orbital eccentricity of a binary system. First, we calculate the fraction of total sources that are detected in each detector above a certain threshold SNR. In Sec.~\ref{sec:resolved binaries}, we quantify the fraction of detected sources that can be resolved. To recall, a binary with $\Delta e_0/e_0<1$ is called a {\it resolved binary}. We show the cumulative distribution of resolved binaries for each eccentricity distribution in Sec.~\ref{sec:distribution of resolved binaries}. In Sec.~\ref{sec:de0vse0}, we investigate the minimum values of eccentricity that can be measured with Voyager, CE, and ET. 

\subsection{Fraction of detected binaries with resolved eccentricity}\label{sec:resolved binaries}
We draw $20000$ BBH samples from mass, spin, eccentricity, and redshift distributions discussed in Sec.~\ref{sec:bbh population}. For Voyager, CE, and ET, we calculate the SNR for all these sources. The detection threshold is set to be $\rho_{\rm th}\geq8$ for all the detectors. Table~\ref{tab:snr} summarizes the fraction of sources that are detected in Voyager, CE, and ET with $\rho_{\rm th}\geq8$. For each detector, we perform five independent realisations of the simulation and quote the median value of the fraction of detected binaries. Voyager detects $33\%$ of the total sources, while CE and ET detect $100\%$ and $97\%$ of the total sources with $\rho_{\rm th}\geq8$, respectively. The sources that do not pass the detection threshold are discarded from further analysis.

For detected binaries, we calculate $1\sigma$ errors on binary parameters using the Fisher information matrix. Table~\ref{tab:fraction} shows the fraction of detected binaries for which eccentricity can be resolved. A source with $\Delta e_0/e_0<1$ can be called eccentric with $\gtrsim 68\%$ confidence. The value quoted is the median value of the five independent realisations. Among all eccentricity distributions, $\text{Loguniform}(10^{-4}, 0.2)$ represents the maximum number of resolved sources for all three detectors. This is because $\text{Loguniform}(10^{-4}, 0.2)$ has the maximum number of sources with higher eccentricity. For this eccentricity distribution, Voyager can confidently measure non-zero eccentricity for $14\%$ of the detected sources. CE can resolve $36\%$ of the sources, while ET can resolve eccentricity for $55\%$ of the sources. The number of resolved binaries in ET is $\sim 1.5$ times larger than CE and $\sim 4$ times larger than Voyager. This is due to the longer inspiral of the GW signal and better low-frequency sensitivity of ET. The effect of the lower cut-off frequency of ET on the fraction of resolved binaries is discussed in Appendix~\ref{sec:appendix}.

For $\text{Loguniform}(10^{-7}, 0.2)$, Voyager can resolve $8\%$ of the detected sources, while CE and ET resolve $20\%$ and $30\%$ of the detected sources, respectively. Lowering the lower limit of eccentricity distribution to $e_0=10^{-7}$ reduces the number of sources with higher eccentricity. Therefore, the number of resolved binaries reduces by a factor of $\sim 2$ compared to $\text{Loguniform}(10^{-4}, 0.2)$. ET can still resolve the eccentricity for $30\%$ of the detected sources for $\text{Loguniform}(10^{-7}, 0.2)$.

For $\text{Powerlaw}(10^{-4}, 0.2)$, the number of sources for which the eccentricity can be resolved reduces by a factor of $1.5 \mbox{--} 2$ compared to the $\text{Loguniform}(10^{-4}, 0.2)$ distribution. ET can resolve $37\%$ of the detected sources for $\text{Powerlaw}(10^{-4}, 0.2)$. For $\text{Powerlaw}(10^{-7}, 0.2)$, the number of resolved sources becomes $\sim 4$ times smaller for $\text{Powerlaw}(10^{-4}, 0.2)$ for all detectors. ET will be able to resolve eccentricity for $\sim 9\%$ of the sources for $\text{Powerlaw}(10^{-7}, 0.2)$ distribution. 

For $\text{Zevin}(10^{-7}, 0.2)$, Voyager can confidently distinguish $3\%$ of the detected sources from circular binaries. CE can distinguish $9\%$ of the sources as eccentric, while ET can resolve $13\%$ of the detected sources. In Sec.~\ref{sec:de0vse0}, we investigate which part of the eccentricity distributions dominates the resolved binaries.

Eccentricity distributions based on N-body simulations (including Zevin eccentricity distribution) predict that a reasonable fraction of binaries can retain $e_0 >0.2$ at $10$ Hz GW frequency. A binary with a higher value of eccentricity is easier to be distinguished as eccentric. We do not include sources with $e_0>0.2$ (at $10$ Hz) in our analysis. Hence the fraction of resolved binaries quoted in Table~\ref{tab:fraction} represents a conservative lower limit. 

\begin{figure*}
   \centering
    \begin{subfigure}{\includegraphics[width=0.45\textwidth]{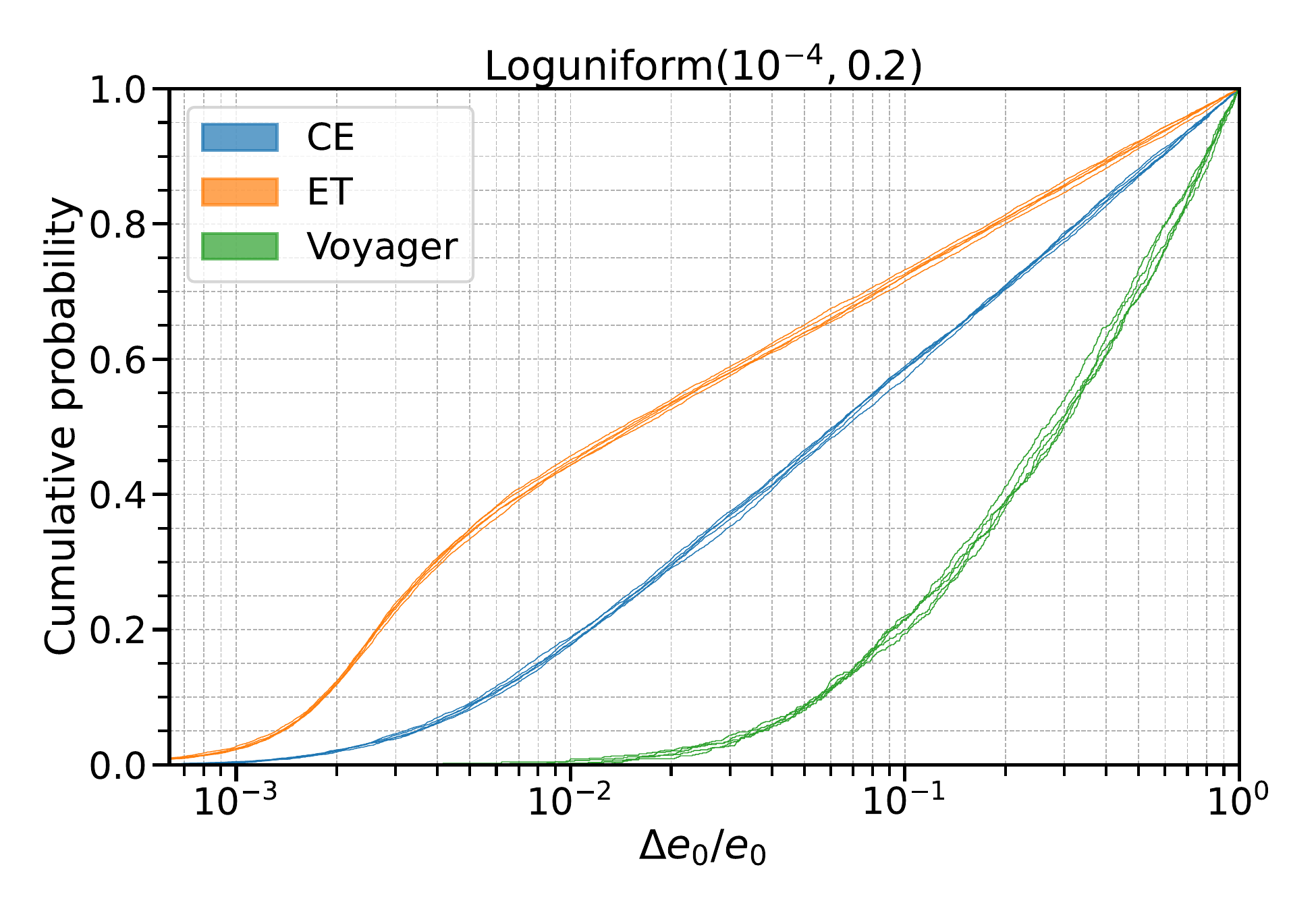}}
    \end{subfigure}
    \vspace{-0.5cm}
    \begin{subfigure}{\includegraphics[width=0.45\textwidth]{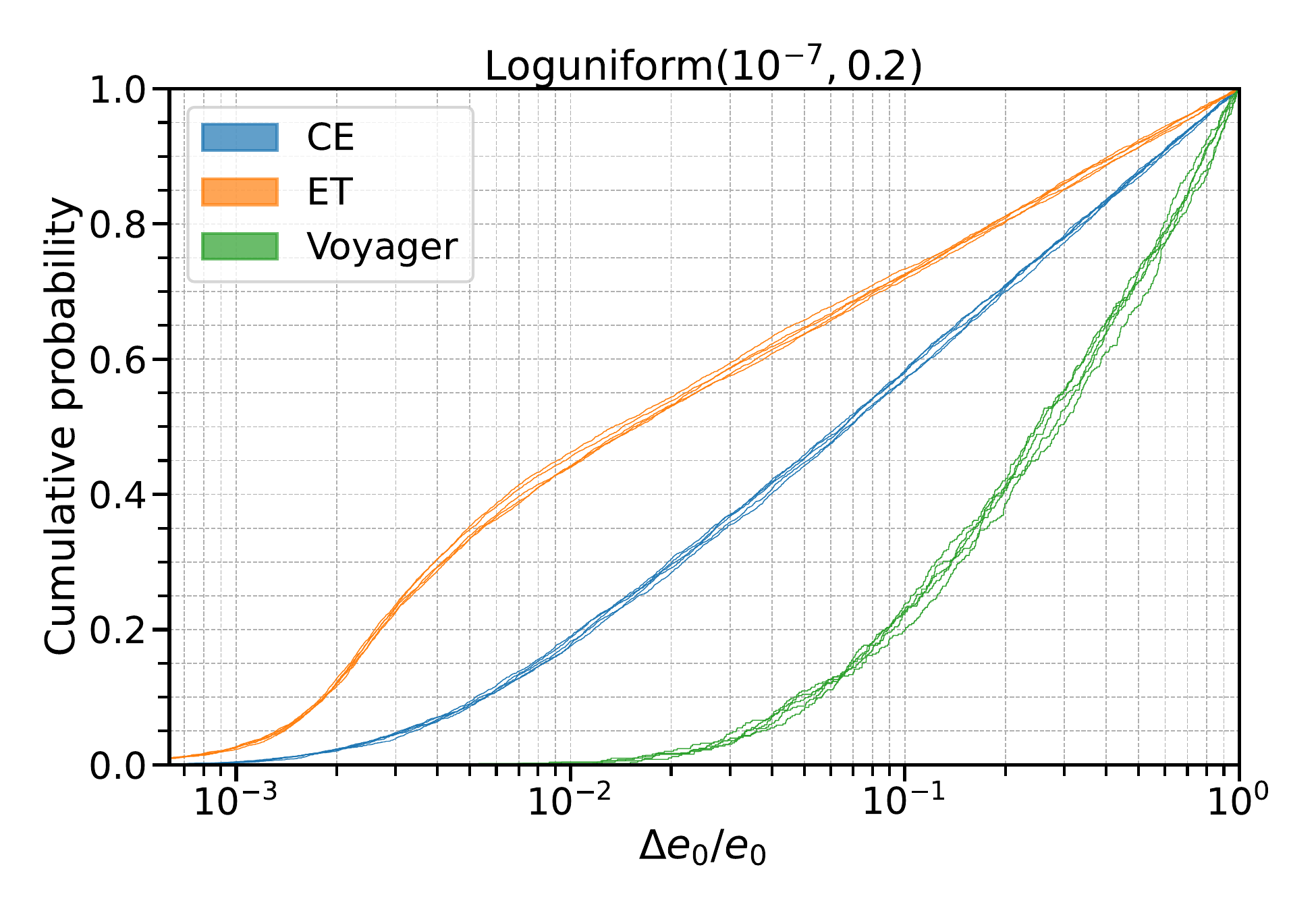}}
    \end{subfigure}
        \vspace{-0.5cm}
        \begin{subfigure}{\includegraphics[width=0.45\textwidth]{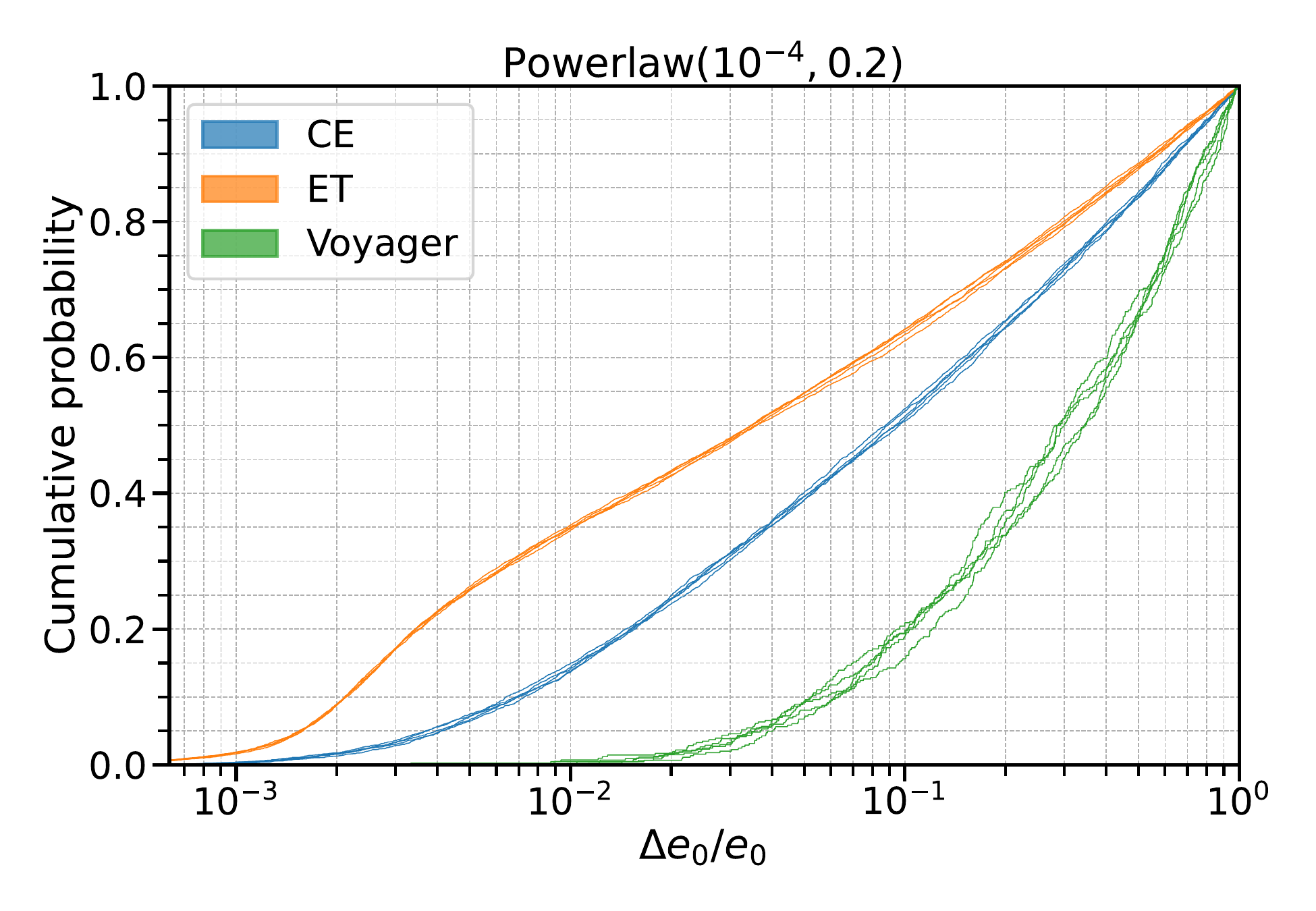}}
    \end{subfigure}
    \begin{subfigure}{\includegraphics[width=0.45\textwidth]{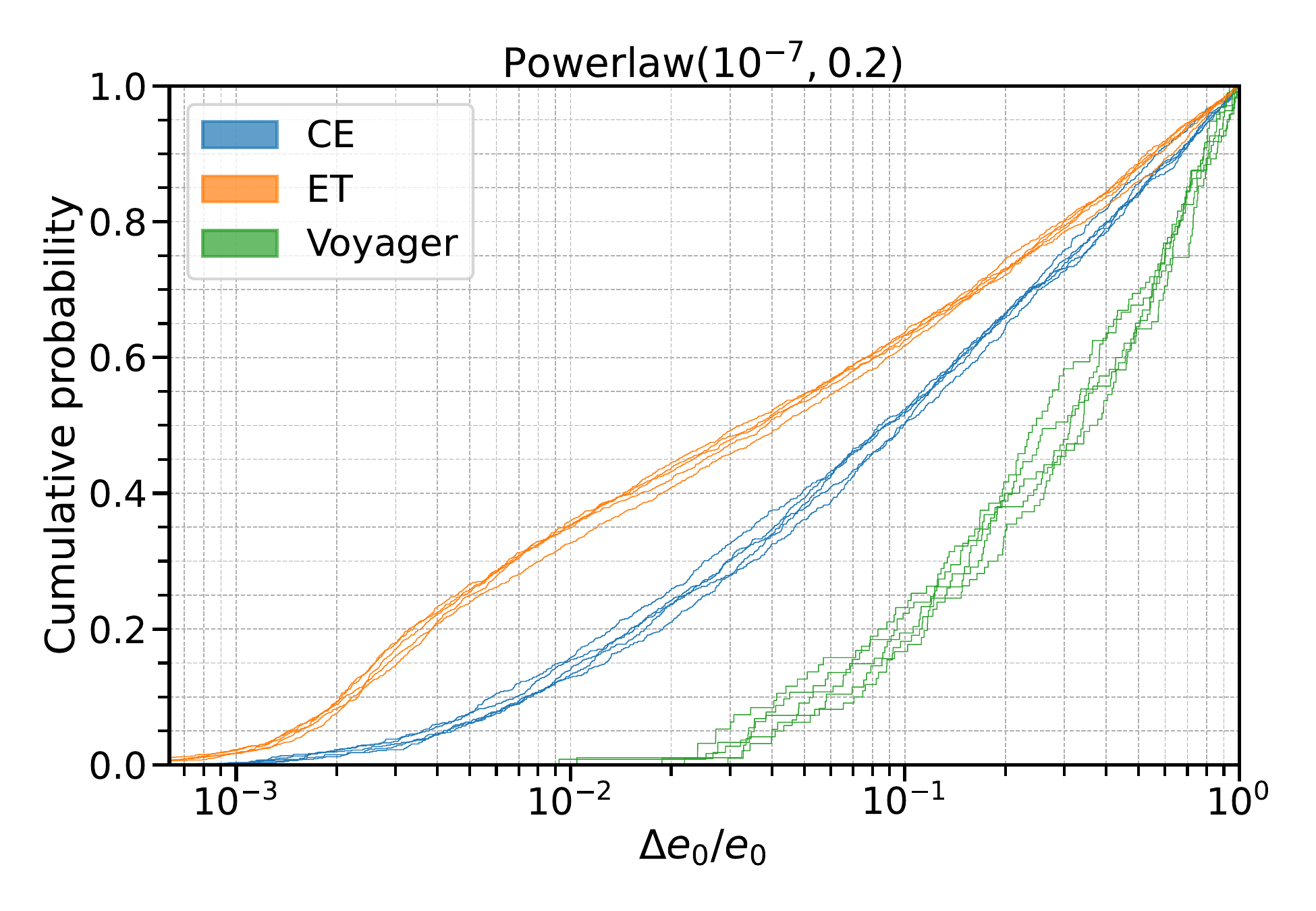}}
    \end{subfigure}
    \begin{subfigure}{\includegraphics[width=0.45\textwidth]{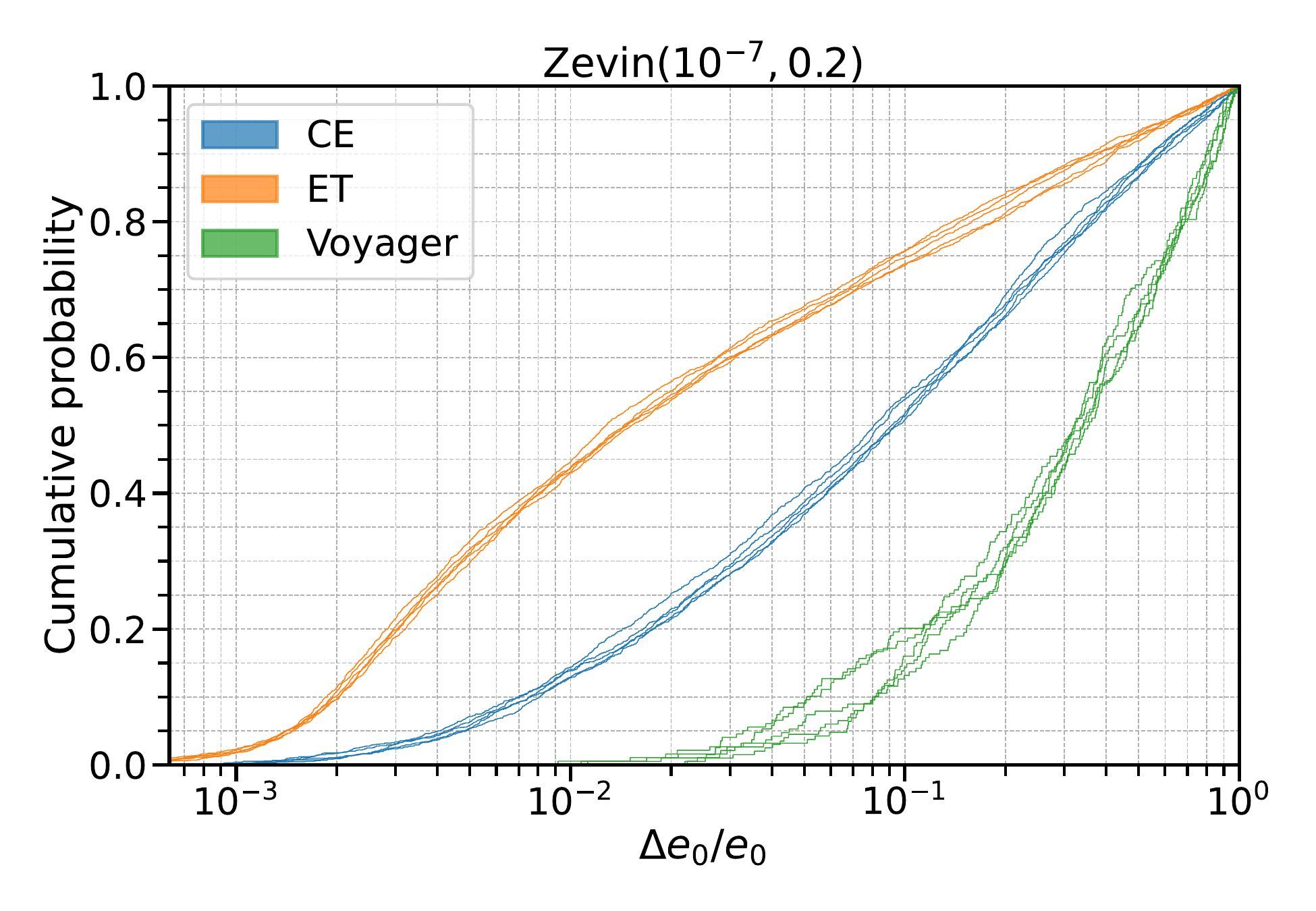}}
    \end{subfigure}
    \caption{The cumulative probability distribution of $\Delta e_0/e_0$ for resolved binaries. Different panels show different eccentricity distributions. Three colours in each panel represent different GW detectors. For each detector, thin lines with the same colour represent five different realisations of the population. ET measures a larger fraction of binaries with smaller errors for all eccentricity distributions. The lower frequency cut-off frequency $(f_{\rm low})$ for Voyager and CE is $5$ Hz. For ET, $f_{\rm low}=1$ Hz.} 
    \label{fig: distribution of errors }
\end{figure*}

\subsection{Fraction of resolved binaries with excellent eccentricity measurement}\label{sec:distribution of resolved binaries}
Having discussed the ability of GW detectors to constrain the fraction of different eccentricity distributions by resolving the eccentricity of binaries in Sec.~\ref{sec:resolved binaries}, in this section, we discuss the cumulative distribution of $\Delta e_0/e_0$ for resolved binaries to get insight into the number of resolved binaries and their corresponding measurement precision. Figure~\ref{fig: distribution of errors } shows the cumulative probability distributions of $\Delta e_0/e_0$ for resolved binaries. Different panels show different eccentricity distributions. Three curves in different colours represent three GW detectors: Voyager, CE, and ET. Each curve consists of five thin curves which represent five independent realisations. 

For all eccentricity distributions, ET measures eccentricity for the larger number of sources with smaller errors followed by CE and Voyager. For $\text{Loguniform}(10^{-4}, 0.2)$, Voyager is able to measure $\sim20\%$ of the resolved binaries with precision better than $10\%$. While CE and ET can measure $\sim60\%$ and $\sim75\%$ of the resolved sources with a precision better than $10\%$, respectively. For $\text{Loguniform}(10^{-7}, 0.2)$, the fraction of resolved sources with $\Delta e_0/e_0 \leq 0.1$ is similar to $\text{Loguniform}(10^{-4}, 0.2)$ distribution. It is important to stress that though the fraction of resolved binaries with $\Delta e_0/e_0 \leq 0.1$ is similar for $\text{Loguniform}(10^{-4}, 0.2)$ and $\text{Loguniform}(10^{-7}, 0.2)$, the number of resolved binaries is less for $\text{Loguniform}(10^{-7}, 0.2)$ compared to $\text{Loguniform}(10^{-4}, 0.2)$. For $\text{Powerlaw}(10^{-4}, 0.2)$, the fractions of resolved sources with precision better than $10\%$ for Voyager, CE, and ET are $\sim 20\%, 50\%$, and $65\%$, respectively. The fractions of resolved binaries are similar for $\text{Loguniform}(10^{-7}, 0.2)$. 

For the $\text{Zevin}(10^{-7}, 0.2)$ distribution, Voyager can measure the eccentricity of $\sim 15\%$ of the resolved sources with accuracy better than $10\%$. CE measures $\sim 52\%$ of the resolved sources with $\Delta e_0/e_0<0.1$, whereas ET measures eccentricity for $\sim 75\%$ of the resolved sources ($13\%$) with better than $10\%$ precision. ET shows remarkable capabilities for measuring the eccentricity of binary black holes. The dependence of eccentricity measurement on the binary's eccentricity is explained in the next Section.

\subsection{Minimum resolvable eccentricity for Voyager, Cosmic Explorer, and Einstein Telescope}\label{sec:de0vse0}
\begin{figure*}
   \centering
    \begin{subfigure}{\includegraphics[width=0.40\textwidth]{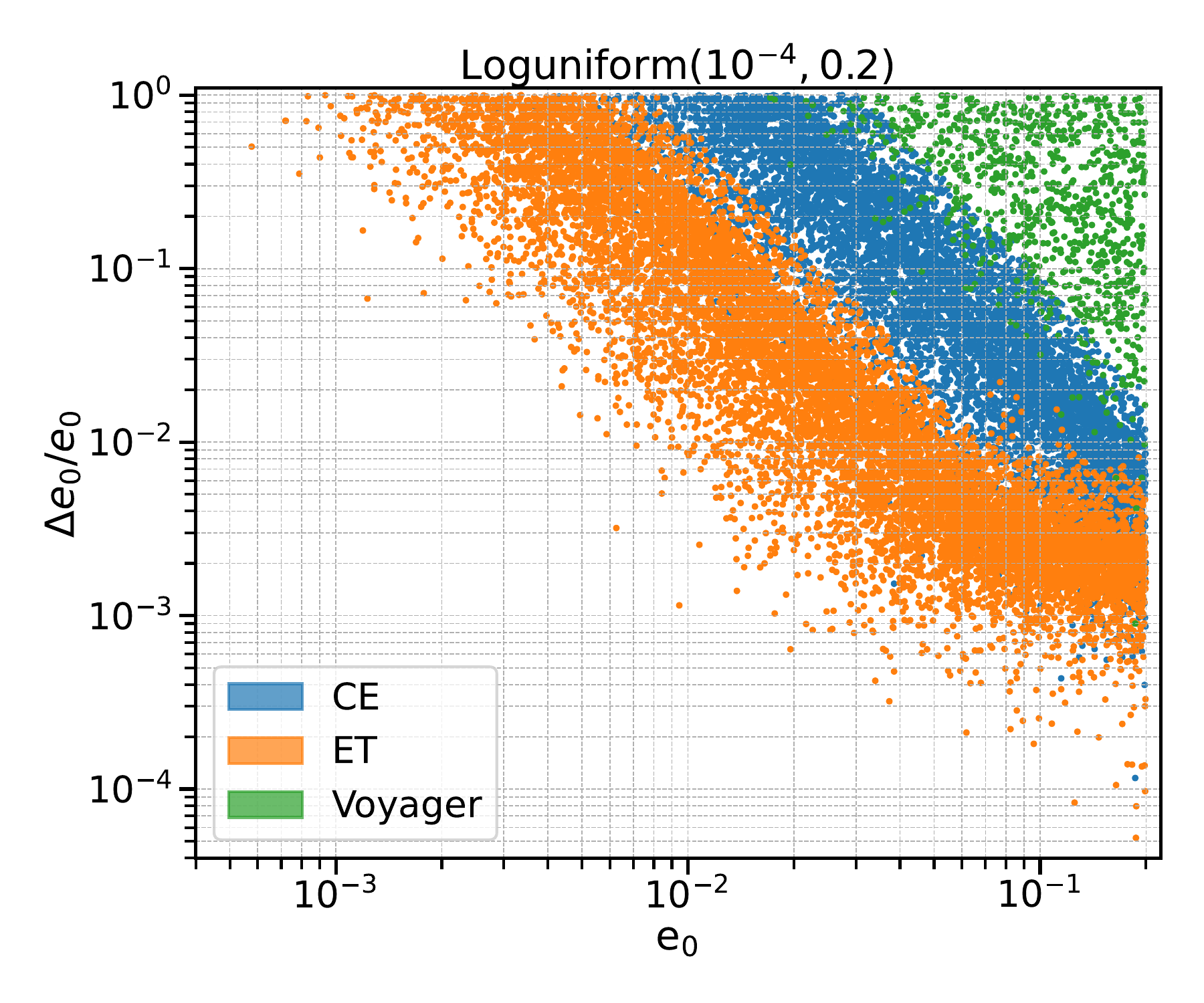}}
    \end{subfigure}
    \vspace{-0.65cm}
    \begin{subfigure}{\includegraphics[width=0.40\textwidth]{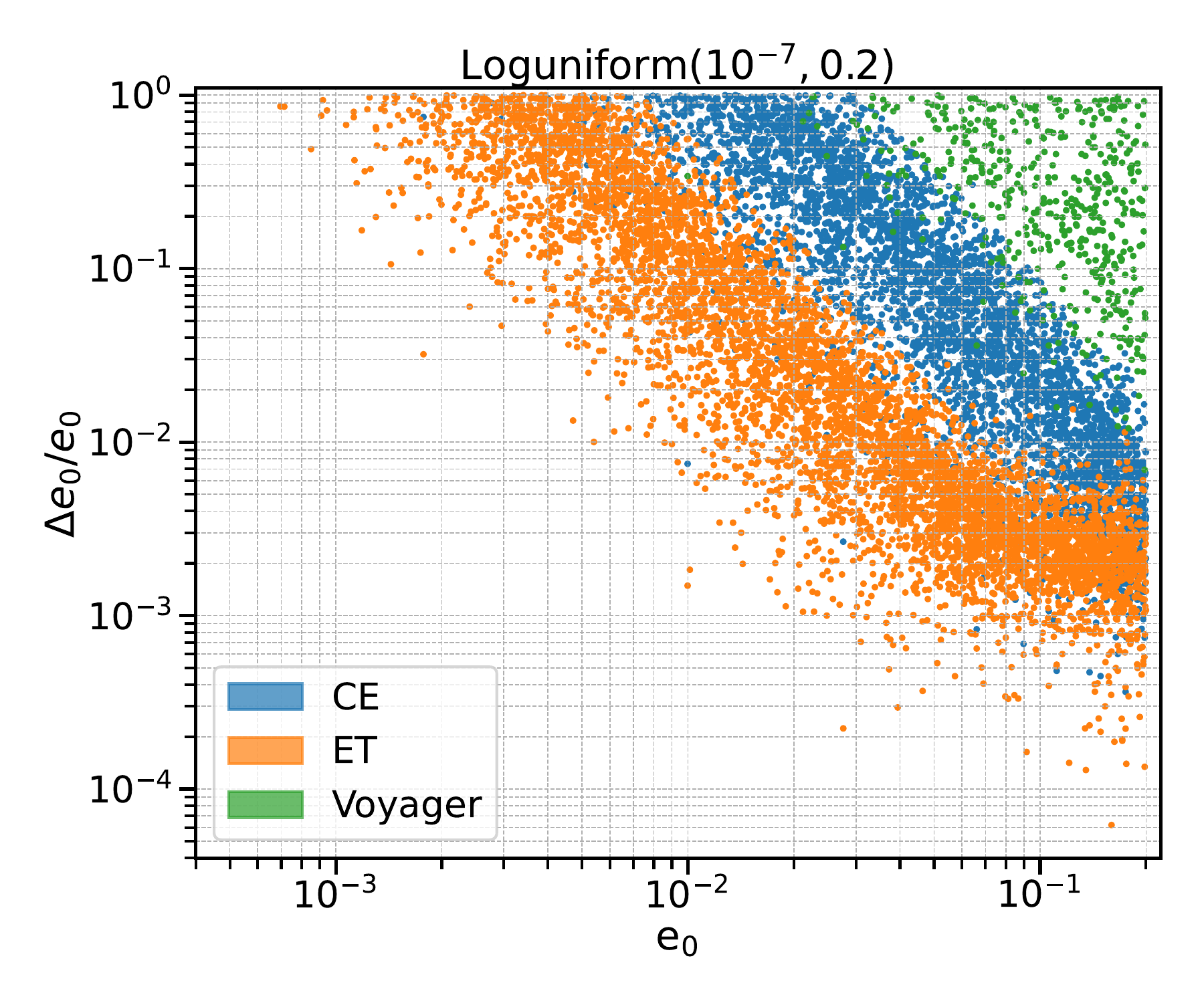}}
    \end{subfigure}
    \begin{subfigure}{\includegraphics[width=0.40\textwidth]{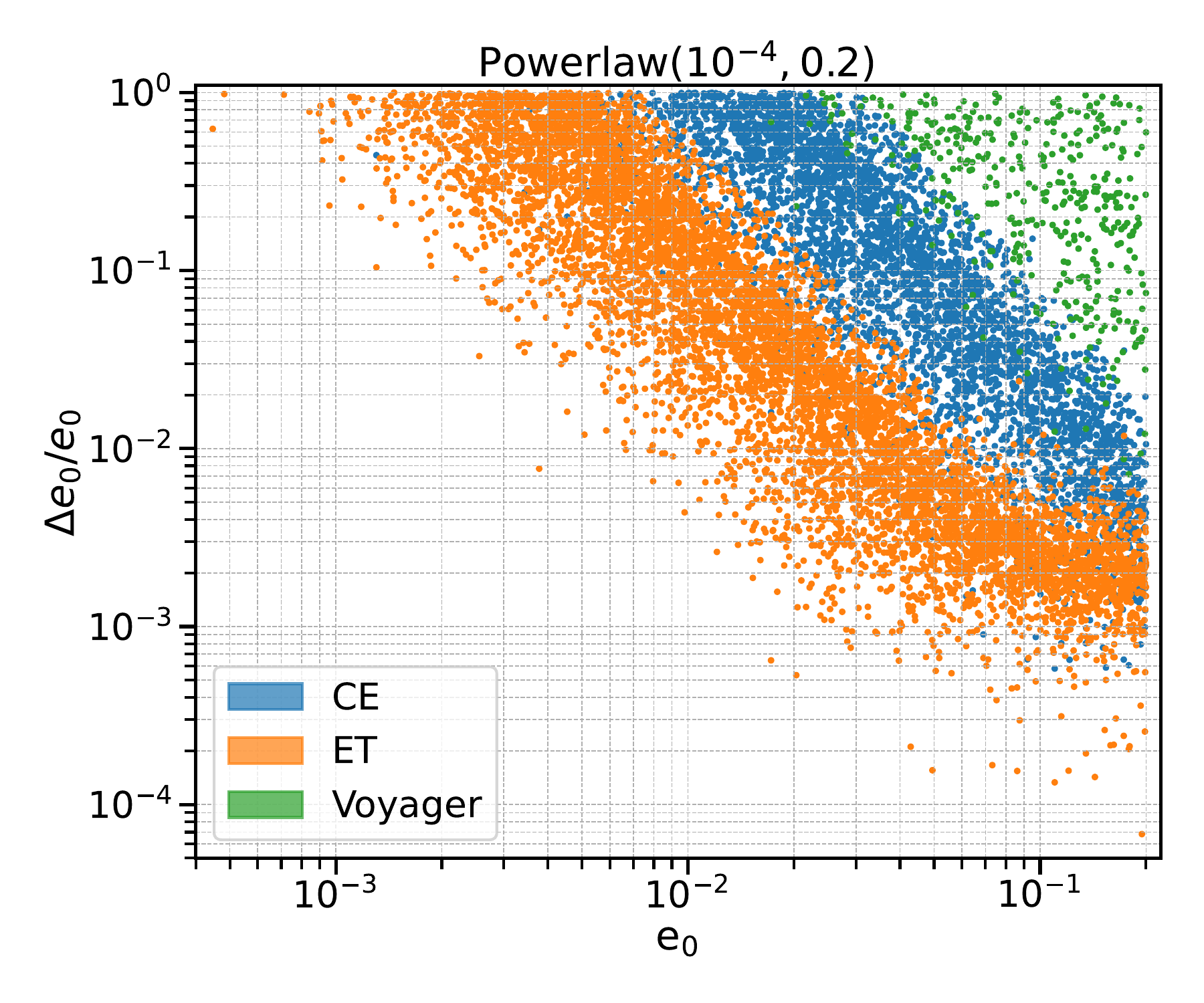}}
    \end{subfigure}
    \vspace{-0.65cm}
    \begin{subfigure}{\includegraphics[width=0.40\textwidth]{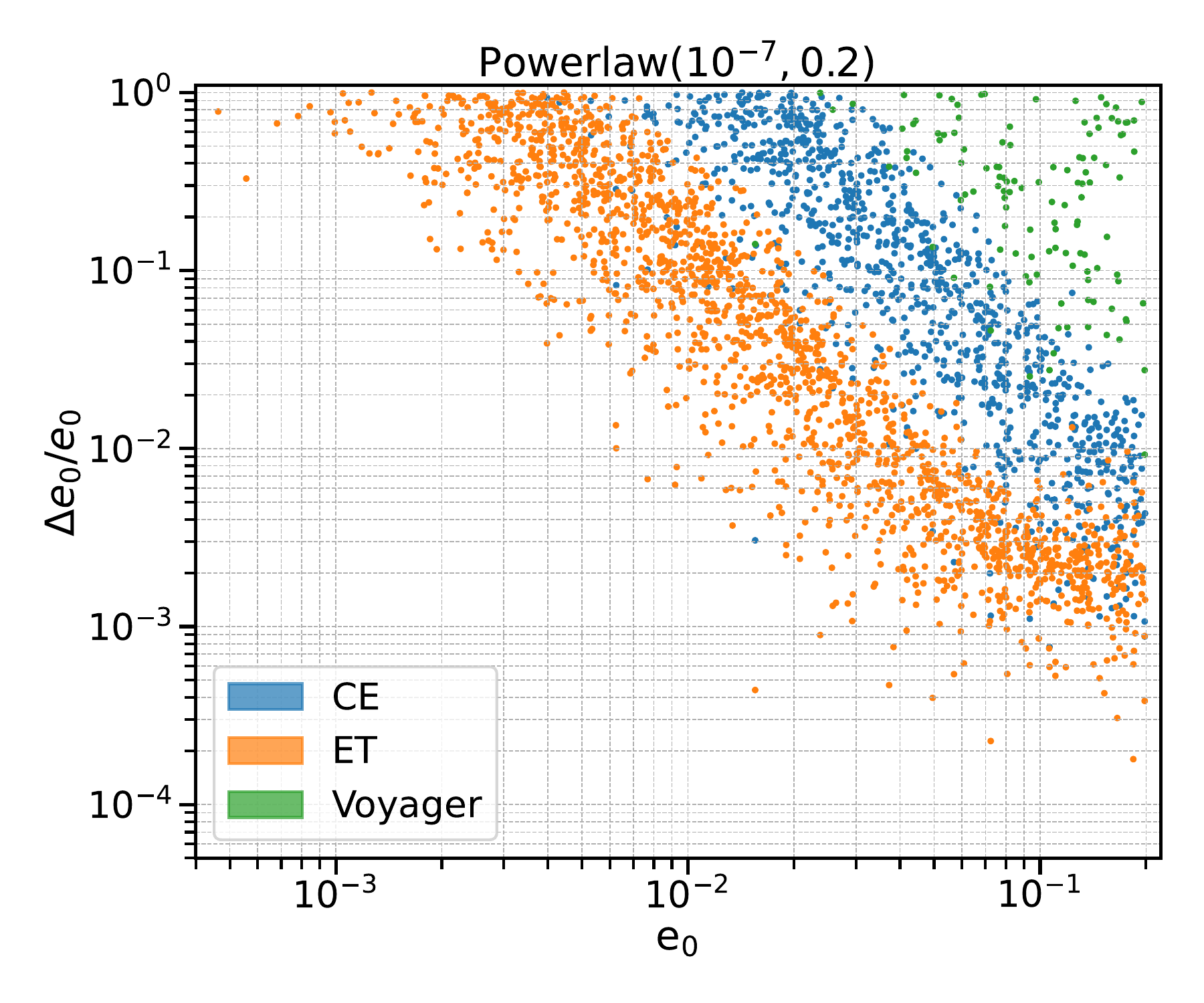}}
    \end{subfigure}
    \begin{subfigure}{\includegraphics[width=0.40\textwidth]{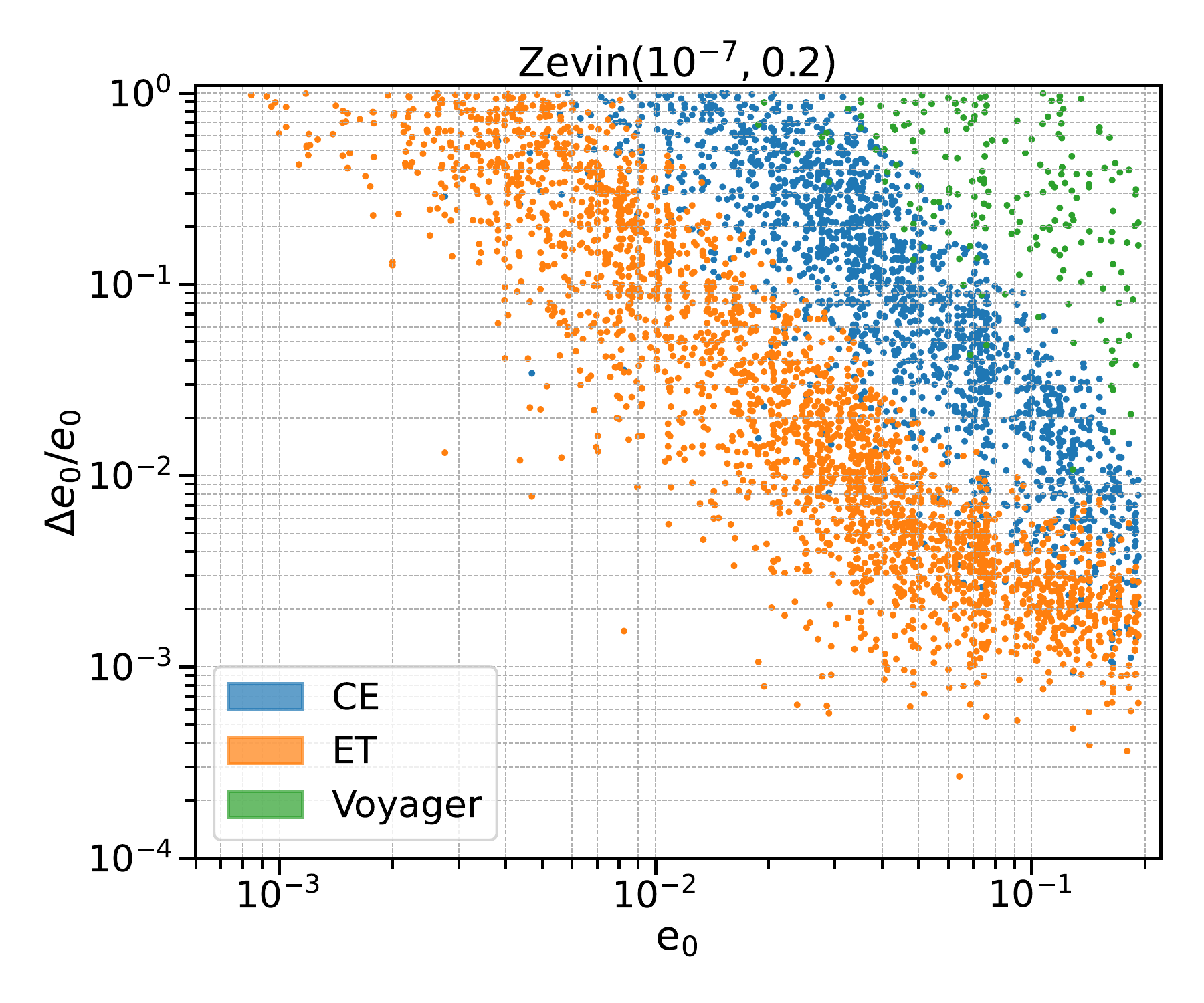}}
    \end{subfigure}
    \caption{Scatter plots of $\Delta e_0/e_0$ with $e_0$ for those sources that have $\Delta e_0/e_0<1$. Each panel shows different eccentricity distributions. The distinct colours represent different detectors. The initial orbital eccentricity $e_0$ on the x-axis is defined at a reference gravitational-wave (dominant mode) frequency of $10$ Hz.} 
    \label{fig: scatter plot}
\end{figure*}
To gain insights into the measured values of eccentricities and their corresponding precision, we plot $\Delta e_0/e_0$ as a function of $e_0$ for resolved binaries in Fig.~\ref{fig: scatter plot} for one of the (randomly chosen) realisations. Figure~\ref{fig: scatter plot} shows the dependence of eccentricity measurement on binary's eccentricity. Five panels represent different eccentricity distributions. In each panel, distinct colours represent different detectors. For all eccentricity distributions, each detector takes a different region in ($e_0$, $\Delta e_0/e_0$) plane with ET measuring more number of sources with better precision followed by CE and Voyager. As expected, the most precise measurement of eccentricity comes from the larger eccentricities. 

From the plots, it is clear that as the eccentricity increases, the $\Delta e_0/e_0$ decreases. For ET, the fractional errors reduce to $\sim 10^{-3}-10^{-2}$ for high values of eccentricities ($e_0>0.1$). For CE, a source with $e_0>0.1$ can be measured with $\Delta e_0/e_0$ $\sim 5\times10^{-3}-0.1$. Voyager can measure eccentricity greater than $0.1$ with $\Delta e_0/e_0$ $\sim0.1-1$. Notice that the number of sources with measured eccentricity is smaller for eccentricity distributions with a wider range $(10^{-7},0.2)$, as expected. 

It can be observed from the figure that it requires $e_0\gtrsim 10^{-3}$ at $10$ Hz GW frequency for ET to confidently distinguish a binary system as eccentric. For CE and Voyager, it requires $e_0\gtrsim 5\times 10^{-3}$ and $e_0\gtrsim 0.02$ (at $10$ Hz GW frequency), respectively, to confidently measure a non-zero eccentricity. The better measurement of eccentricity by ET compared to CE is mainly due to two effects. The first is due to the better sensitivity of ET at low frequencies (below $8$ Hz) compared to CE and the second is because the binary system has a relatively larger value of eccentricity when it enters the frequency band of ET ($1$ Hz). For example, a binary with $e_0=0.05$ at $10$ Hz GW frequency had an eccentricity of $\sim 0.1$ when it entered the CE and Voyager band at $5$ Hz. The binary had an eccentricity of $\sim 0.4$ when entering the ET band at $1$ Hz \footnote{It is worth noting that for few sources the value of $e_0$ can become greater than $0.2$ in the frequency band of the detector (especially for ET). The waveform model becomes less accurate for $e_0>0.2$. However, the number of sources with such high eccentricity is very small compared to the total number of sources.}. Hence ET provides better measurements of eccentricity. 

It is clear that Voyager, CE, and ET can resolve only the {\it high eccentricity} part of the eccentricity distributions. None of the detectors is able to resolve eccentricity $\lesssim 10^{-3}$ at $10$ Hz GW frequency. In Zevin eccentricity distribution, the BBHs formed via GW capture have $e_0 \gtrsim 10^{-3}$ at $10$ Hz GW frequency. Hence next generation GW detectors such as Voyager, CE, and ET will be able to confidently distinguish a BBH system as eccentric that is likely formed via GW capture in globular clusters. However, there are other channels that can produce BBHs with $e_0\gtrsim10^{-3}$ at $10$ Hz. These include gravitational interaction of field triples~\citep{Silsbee:2016djf,Antonini:2017ash, 2018ApJ...863....7R,Michaely:2019aet, Grishin:2021hcp,Raveh:2022ste}, three-body mergers in globular clusters~\citep{PhysRevD.97.103014}, direct capture within galactic nuclei~\citep{OLeary:2008myb}, and binaries formed through secular GW evolution near supermassive BHs~\citep{Antonini:2012ad, 2014ApJ...794..122M, 2018ApJ...856..140H,PhysRevD.98.083028, 2018ApJ...856..140H}. Though next generation GW detectors would be able to identify a BBH system as eccentric formed through these channels, distinguishing these formation scenarios from each other may be more challenging. 

\section{Summary and outlook}\label{sec:summary}
In this work, we have studied the ability of next generation GW detectors such as Voyager, CE, and ET to measure the eccentricity of BBHs. The key goal was to investigate the fraction of detected binaries that can be confidently distinguished as eccentric by measuring a non-zero value of eccentricity. The precise measurement of eccentricity can be used to distinguish different formation channels as the eccentricity of a binary system depends on its formation history.  

We find that for Zevin eccentricity distribution, Voyager can resolve the eccentricity for $\sim 3\%$ of the detected BBHs. CE can resolve eccentricities for $\sim 9\%$ of the detected BBHs, whereas ET can resolve the eccentricity of $\sim 13\%$ of the detected BBHs. Moreover, a subpopulation of resolvable eccentric binary systems will have an extremely precise measurement of eccentricity. For Voyager, $(15\mbox{--}20)\%$ of the resolved binaries will have eccentricity measurement better than $10\%$, while for CE and ET, $(50\mbox{--}60)\%$ and $(65\mbox{--}75)\%$ of the resolved binaries will have $\Delta e_0/ e_0 \lesssim10\%$, respectively. ET's low-frequency sensitivity plays a crucial role in the measurement of eccentricity.

Another important aspect of this study is the determination of the minimum value of eccentricity that is required to confidently identify a binary system as eccentric. We find that it requires eccentricities $\gtrsim 0.02$ (at $10$ Hz GW frequency) for Voyager to identify a binary as eccentric. For CE and ET, the minimum eccentricities that can be measured to distinguish a binary as eccentric are $\gtrsim 5\times 10^{-3}$ and $\gtrsim 10^{-3}$ at $10$ Hz GW frequency, respectively. In summary, next generation GW detectors would be able to confidently distinguish BBH systems with $e_0\gtrsim 10^{-3}$ (at $10$ Hz GW frequency) from circular binaries. 

Dynamically formed binaries are expected to have isotropic spin distribution. Our waveform model does not account for the precessional effect. Moreover, {\tt TaylorF2Ecc} waveform model does not take into account the higher modes due to eccentricity. It would be interesting to study the effect of precession and higher modes (due to eccentricity) on eccentricity measurement. The ability of GW detectors to reconstruct the actual eccentricity distribution can be explored in a future study.
 
\section*{Acknowledgements}
P.~S. is grateful to K.~G.~Arun for valuable discussions and comments on the draft. We thank Michael Zevin for providing data to reproduce the eccentricity distribution of \cite{Zevin:2021rtf}. It is a pleasure to thank Anuradha Gupta, Sajad A. Bhat, Parthapratim Mahapatra, Poulami Dutta Roy, Nathan Johnson-McDaniel, and B.~S.~Sathyaprakash for useful discussions and/or comments on the draft. P.~S. acknowledges partial support from the Infosys Foundation. The author is grateful for computational resources provided by the LIGO Laboratory and supported by the National Science Foundation Grants PHY-0757058 and PHY-0823459. This study made use of the following software packages:
{\tt Astropy}~\citep{2013A&A...558A..33A,2018AJ....156..123A}, {\tt Scipy}~\citep{2020NatMe..17..261V},
{\tt NumPy}~\citep{2020Natur.585..357H}, {\tt Matplotlib}~\citep{4160265}, {\tt Seaborn}~\citep{Waskom2021}, {\tt jupyter}~\citep{soton403913}, {\tt pandas}~\citep{mckinney-proc-scipy-2010}.

\section*{DATA AVAILABILITY}
The data supporting the findings of this article will be made available upon reasonable request to the corresponding author.








\bibliographystyle{mnras}
\bibliography{ref}

\appendix

\section{The effect of Einstein Telescope lower cut-off frequency on the fraction of resolved binaries}\label{sec:appendix}

\begin{table*}
\centering
\renewcommand{\arraystretch}{1.2}
\begin{tabular}{cccc}
\toprule
\toprule
&\multicolumn{3}{c} {Fraction (in \%) of detected binaries with $\Delta e_0/e_0<1$}   \\
\midrule
 Eccentricity distribution & \textbf{ET} $(f_{\rm low}=1 \text{Hz})$  & \textbf{ET} $(f_{\rm low}=3 \text{Hz})$ & \textbf{ET} $(f_{\rm low}=5 \text{Hz})$ \\
\midrule
\textbf{Loguniform$(\bm{10^{-4},0.2)}$}  & $55$ & $48$  & $40$ \\
\textbf{Loguniform}$(\bm{10^{-7},0.2)}$ & $30$ & $27$ & $22$ \\
\textbf{Powerlaw}$\bm{(10^{-4},0.2)}$  & $37$ & $30$ & $23$ \\
\textbf{Powerlaw}$\bm{(10^{-7},0.2)}$  & $9$ & $7$ &$6$\\
\textbf{Zevin}$\bm{(10^{-7},0.2)}$ & $13$ & $12$ & $10$\\
\bottomrule
\bottomrule
\end{tabular}
\caption{The impact of ET's lower cut-off frequency ($f_{\rm low}$) on the fraction of resolved binaries. The simulation is repeated five independent times and the median value is quoted. The $f_{\rm low}$ plays a crucial role in the eccentricity measurement.}
\label{tab:ET lower freq}
\end{table*}

\begin{figure*}
   \centering
    \begin{subfigure}{\includegraphics[width=0.32\textwidth]{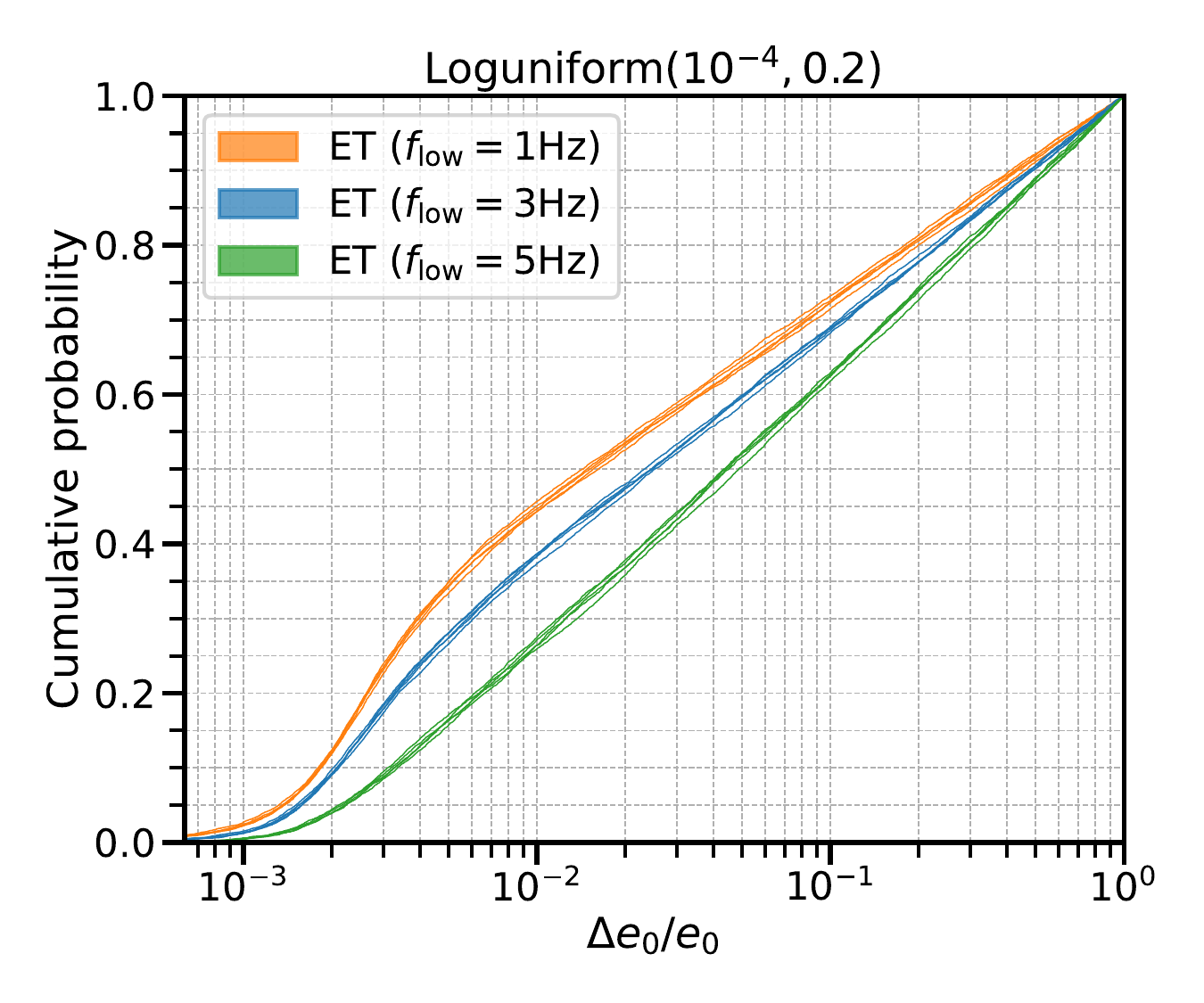}}
    \end{subfigure}
     \begin{subfigure}{\includegraphics[width=0.32\textwidth]{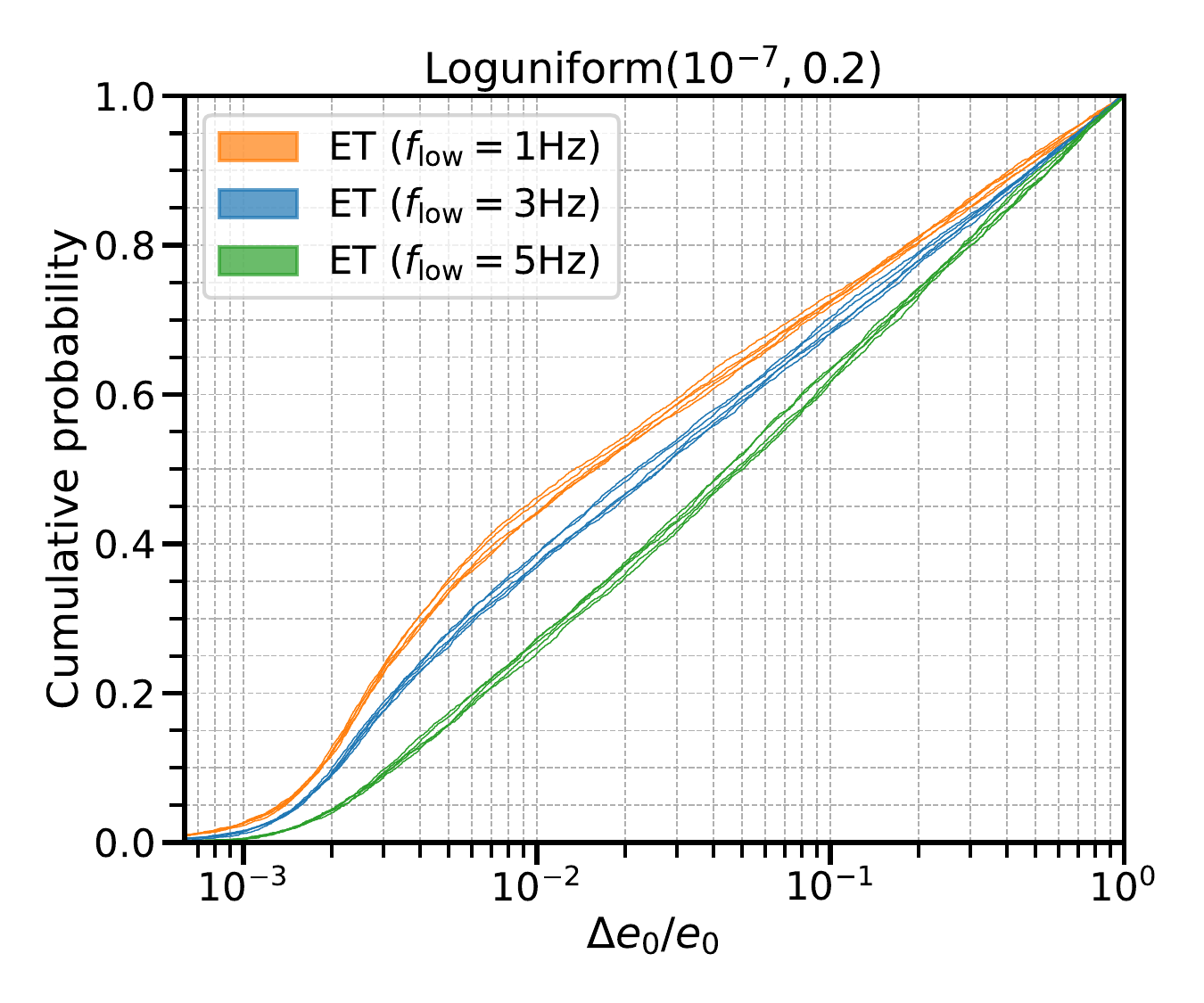}}
    \end{subfigure}
    \begin{subfigure}{\includegraphics[width=0.32\textwidth]{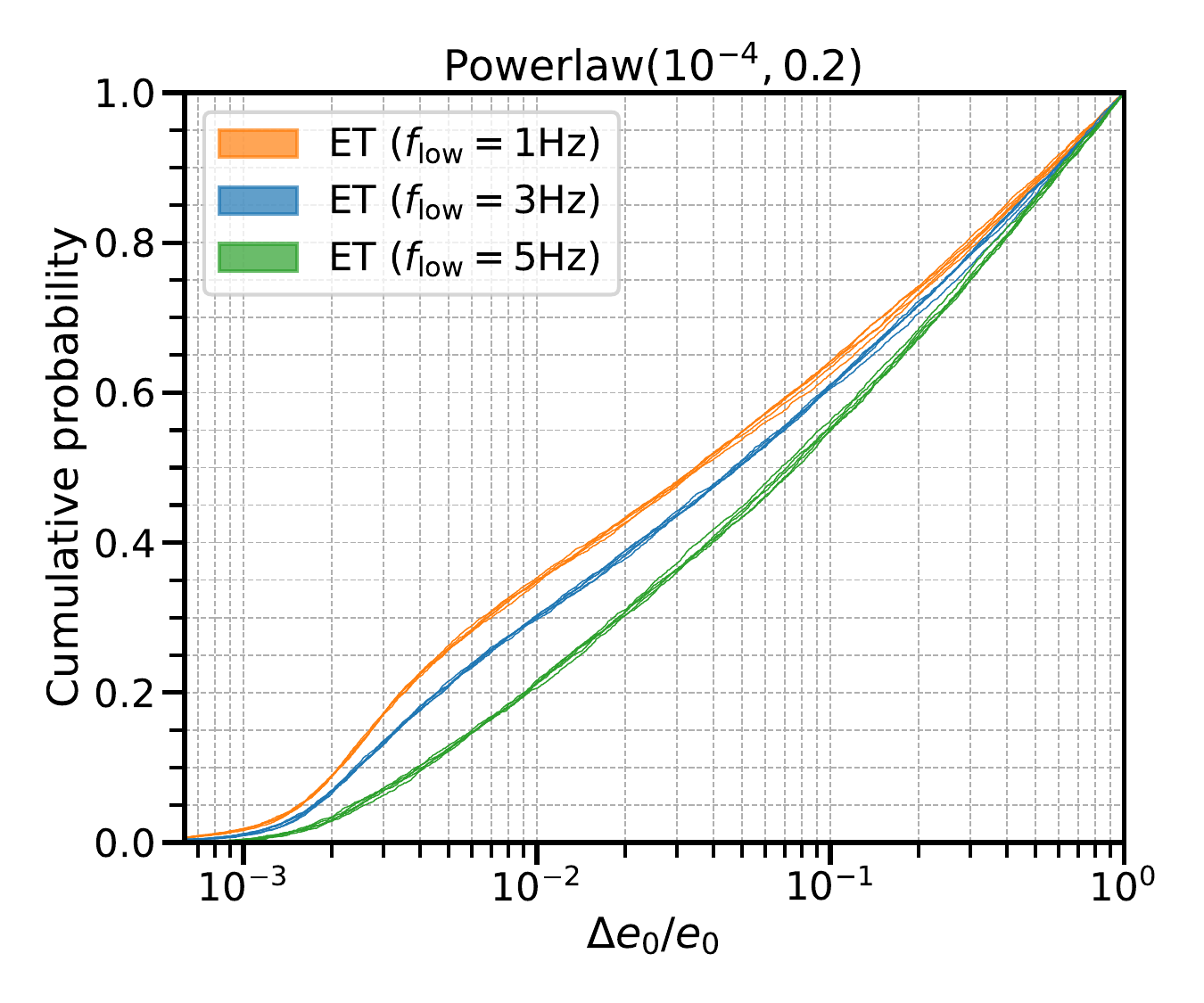}}
    \end{subfigure}
    \begin{subfigure}{\includegraphics[width=0.32\textwidth]{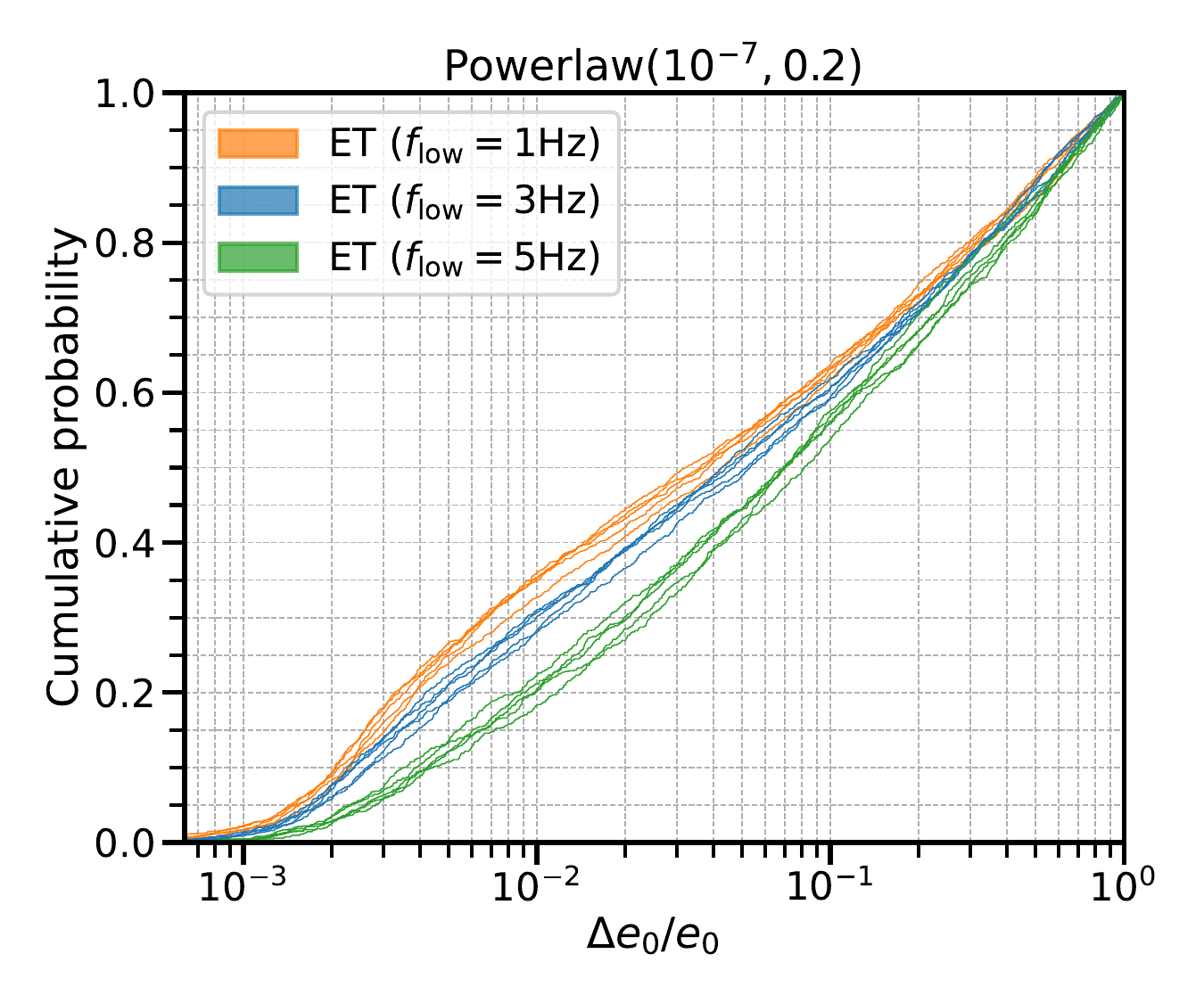}}
    \end{subfigure}
    \begin{subfigure}{\includegraphics[width=0.32\textwidth]{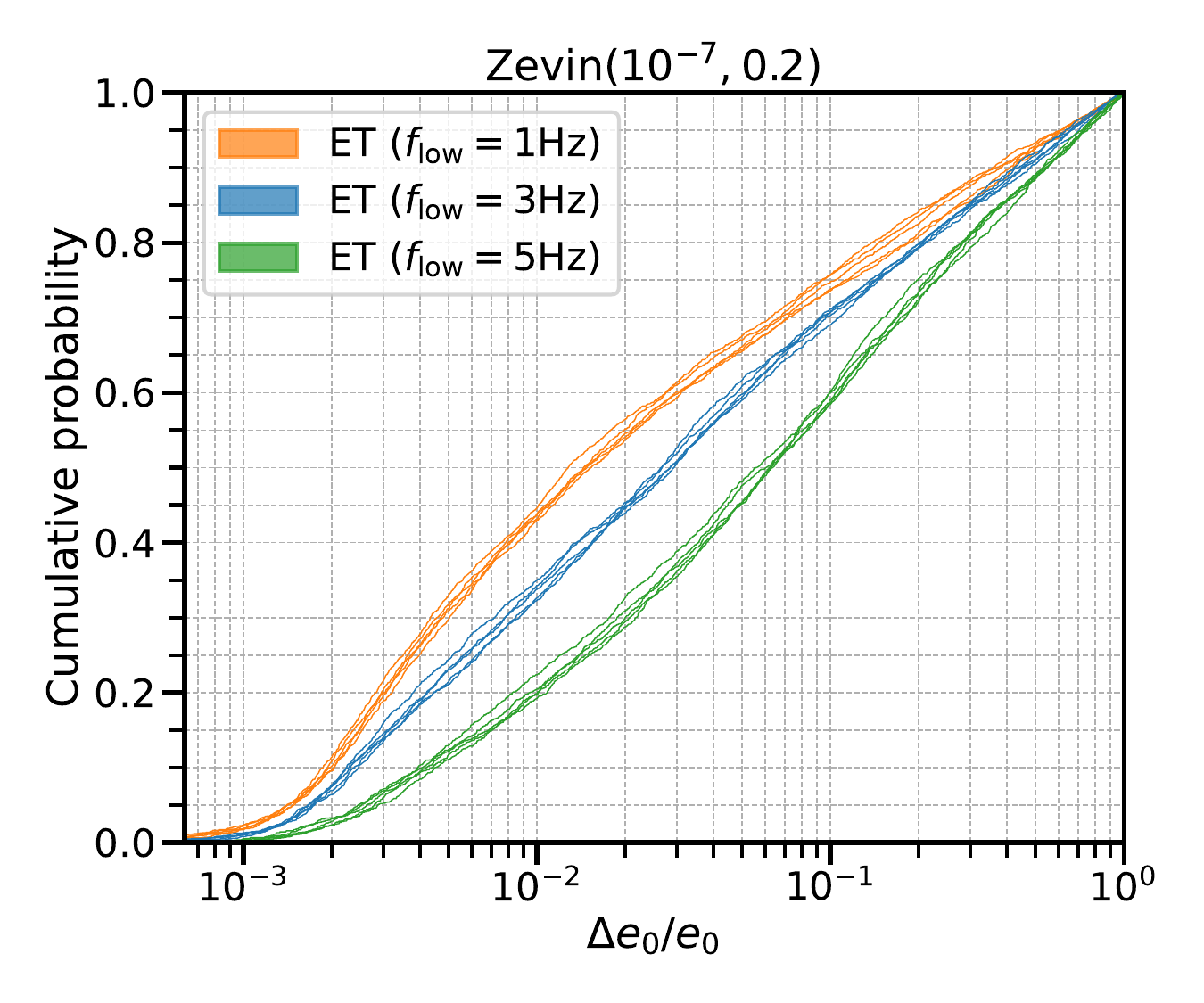}}
    \end{subfigure}
    \caption{The effect of ET's lower cut-off frequency on the cumulative probability distribution. The number of sources with smaller errors reduces as the $f_{\rm low}$ increases from $1$ Hz to $3$ Hz and $5$ Hz.} 
    \label{fig:ET lower freq}
\end{figure*}

Due to the seismic noise, it might be hard to achieve sensitivity as low as $1$ Hz for ET. Here, we study the impact of lower frequency cut-off of ET on the eccentricity measurement. Table~\ref{tab:ET lower freq} shows the fraction of detected binaries with resolvable eccentricity ($\Delta e_0/e_0 <1$) for three frequency cut-offs: $f_{\rm low}=1$ Hz, $f_{\rm low}=3$ Hz, and $f_{\rm low}=5$ Hz. The change of lower frequency cut-off has a significant effect on the fraction of resolved binaries. 

For $\text{Loguniform}(10^{-4}, 0.2)$, increasing the $f_{\rm low}$ from $1$ Hz to $3$ Hz, reduces the fraction of resolved binaries from $55\%$ to $48\%$ (a $\sim 12\%$ change). Further, increasing $f_{\rm low}$ to $5$ Hz, reduces the fraction to $40\%$ ($\sim 27\%$ change). Notice that for $\text{Loguniform}(10^{-4}, 0.2)$, CE ($f_{\rm low}=5$ Hz) can measure eccentricity for $35\%$ of the detected sources. ET at $f_{\rm low}=5$ Hz can resolve a larger fraction of eccentricity distribution compared to CE. This is because of the better low-frequency sensitivity of ET (below $8$ Hz) compared to CE. 

For $\text{Loguniform}(10^{-7}, 0.2)$, there is a $\sim 10\%$ and $\sim 26\%$ reduction in the fraction of resolved sources when $f_{\rm low}$ is increased from $1$ Hz to $3$ Hz and $5$ Hz, respectively. Similar kind of trends can be seen for  $\text{Powerlaw}(10^{-4}, 0.2)$ and $\text{Powerlaw}(10^{-7}, 0.2)$ distributions. For Zevin eccentricity distribution, there is a $7\%$ and $23\%$ reduction in the fraction of resolved sources, when $f_{\rm low}$ is increased from $1$ Hz to $3$ Hz and $5$ Hz, respectively. 

Figure~\ref{fig:ET lower freq} shows the effect of $f_{\rm low}$ on the cumulative probability distribution of $\Delta e_0/e_0 <1$. It can be seen that the number of sources with smaller errors reduces as the $f_{\rm low}$ increases from $1$ Hz to $3$ Hz and $5$ Hz. Hence, the lower cut-off frequency of ET plays a critical role in the measurement of low-frequency effects such as eccentricity.

\bsp	
\label{lastpage}

\end{document}